\documentstyle[pra,aps,twocolumn,psfig,floats,amssymb]{revtex}

\def\be{ \begin{equation} }
\def\ee{ \end{equation} }
\def\ba{ \begin{array} }
\def\ea{ \end{array} }
\def\bea{ \begin{eqnarray} }
\def\eea{ \end{eqnarray} }
\def\bml{ \begin{mathletters} }
\def\eml{ \end{mathletters} }
\def\bmla{ \bml \bea }
\def\emla{ \eea \eml }

\def\H{{\sf H}}
\def\pump{\Omega_P}
\def\Stokes{\Omega_S}
\def\Omeg0{\Omega_0} % peak R.f.
\def\OmRMS{\Omega} % root-mean-square R.f. 
\def\mixangle{\alpha} % mixing angle
\def\th{\vartheta}
\def\NAC{\dot{\th}} % non-adiabatic coupling
\def\Tmax{t_0} % max of non-adb. coupling
\def\ATS{\Phi_T} % trapped state
\def\phaseP{\phi_P} % pump phase
\def\phaseS{\phi_S} % Stokes phase
\def\phase{\phi} % the superposition phase
\def\angleD{\varphi} % rotation angle for nonzero detuning

\begin{document}
\draft
\wideabs{
\author{N. V. Vitanov$^{1,a}$
	K.-A. Suominen$^{1,b}$, and
	B. W. Shore$^2$}
\address{
$^1$Helsinki Institute of Physics, PL 9,
 FIN-00014 Helsingin yliopisto, Finland\\
$^2$Lawrence Livermore National Laboratory,
 Livermore, CA 94550, USA}
\title{Creation of coherent atomic superpositions by fractional STIRAP}
\date{\today }
\maketitle
\begin{abstract}
We discuss a simple scheme for preparing atoms and molecules in an arbitrary
preselected coherent superposition of quantum states.
The technique, which we call fractional stimulated Raman adiabatic passage
({\it f-STIRAP}), is based upon (incomplete) adiabatic population transfer
between an initial state $\psi_1$ and state $\psi_3$ through an intermediate
state $\psi_2$.
As in STIRAP, the Stokes pulse arrives before the pump pulse, but unlike
STIRAP, the two pulses terminate simultaneously
while maintaining a constant ratio of amplitudes.
The independence of f-STIRAP
from details of pulse shape and pulse area
makes it the analog of conventional STIRAP
in the creation of coherent superpositions of states.
We suggest a smooth realization of f-STIRAP which requires only two laser
pulses (which can be derived from a single laser) and at the same time ensures
the automatic fulfillment of the asymptotic conditions at early and late times.
Furthermore, we provide simple analytic estimates of the robustness of
f-STIRAP against variations in the pulse intensity, the pulse delay,
and the intermediate-state detuning, and discuss its possible
extension to multistate systems.
\end{abstract}
\pacs{PACS numbers: 32.80.Bx, 33.80.Be, 42.50.Hz}
}

%======================================================================

\section{Introduction}

\label{Sec-introduction}

Atoms and molecules prepared in a well defined preselected initial state have
applications in many branches of atomic, molecular and optical physics.
There are various techniques available for preparing
a {\it single} initial state \cite{Shore90}.
Optical pumping is the standard method when the desired state is the $m=-J$
or $m=+J$ sublevel of a ground state with a total angular momentum of $J$.
$\pi $-pulse and chirped-pulse techniques are used to prepare atoms and
molecules in a particular excited state, accessible via an electric-dipole
single-photon transition.
A state, accessible via a two-photon transition (which can be an
excited, metastable, or another ground state) can be populated by using
generalized $\pi $-pulses, chirped pulses, or the robust and efficient
technique of stimulated Raman adiabatic passage (STIRAP) \cite
{Kuklinski89,Gaubatz90,Bergmann95,Bergmann98}.
A state accessible via a multiphoton transition can be populated by means of
generalizations of the above techniques.
Some of these techniques are robust against moderate variations in the
interaction parameters, as the chirped-pulse method and STIRAP,
while others, such as the $\pi$-pulse method, are not.

It is considerably more difficult to prepare atoms and molecules in a
{\it preselected coherent superposition} of states.
Creating an initial atomic coherence is essential for
such effects as
dark resonances \cite{Alzetta76,Arimondo76,Alzetta79,Gray79},
subrecoil laser cooling \cite{Aspect88,Lawall94LC,Lawall95,Cohen-Tannoudji90},
electromagnetically induced transparency \cite{Hakuta91,Boller91,Field91},
light amplification without inversion
\cite{Kocharovskaya88,Kocharovskaya92,Nottelmann93},
refraction index enhancement without absorption \cite{Scully91},
harmonic generation \cite{Gauthey95,Watson96}, and
quantum information \cite{Steane98}.
Modifications of the above techniques can still be used but they are
generally very sensitive to small variations in the interaction parameters,
such as pulse areas and detunings.
In this paper, we discuss a technique, based on ideas similar to those of
STIRAP, which guarantees the creation of any desired coherent superposition
of two states.
The technique, which we call {\it fractional STIRAP} (f-STIRAP), is based
upon adiabatic population transfer between the initially populated state
$\psi_1$ and state $\psi_3$ through an intermediate state $\psi_2$.
It requires a two-photon resonance between states $\psi_1$ and $\psi_3$ and
uses two laser pulses, a pump pulse $\pump$, linking states $\psi_1$ and
$\psi_2$, and a Stokes pulse $\Stokes$, linking states $\psi_2$ and $\psi_3$.
As in STIRAP, the Stokes pulse arrives before the pump pulse, but unlike
STIRAP, where the Stokes pulse vanishes first, here the two pulses vanish
simultaneously.
This ``incompleted STIRAP'' evolution provides the possibility of ending with
the population residing in both states $\psi_1$ and $\psi_3$, rather than
being transferred entirely to state $\psi_3$, as in STIRAP.
Moreover, since the population transfer is carried out through an adiabatic
state which is a linear superposition of states $\psi_1$ and $\psi_3$ only
(often referred to as the trapped or dark state), state $\psi_2$ remains
unpopulated, even transiently, and hence its properties, including
spontaneous decay, do not affect the process.
This guarantees the coherence of the created final superposition of states.
The f-STIRAP has all the advantages that STIRAP has in population transfer
to a single state, regarding robustness, efficiency and simplicity, and can
be considered as its analog in creating coherent superpositions of states.

The idea of interrupted STIRAP was proposed for the first time by
Marte, Zoller and Hall \cite{Marte91} as a way to create an atomic beam
splitter, which is an essential part of an atomic interferometer
\cite{Marte91,Lawall94BS,Goldner94,Weitz94PRA,Weitz94PRL,Weiss94,%
Featonby95,Theuer98}.
This idea was later discussed by Lawall and Prentiss \cite{Lawall94BS}
and analyzed in more detail by Weitz, Young and Chu \cite{Weitz94PRA},
who have demonstrated it experimentally \cite{Weitz94PRL}, achieving the
interruption of STIRAP evolution by simultaneously and abruptly turning
to zero the intensities of both the pump and the Stokes fields.
In the present paper, we discuss the potential of f-STIRAP for preparing
{\it coherent superpositions} of states.
This aspect is slightly different because it puts a particular emphasize
on the {\it robustness} of the process.
This is so because variations in the parameters of the created superposition
(the populations and the relative phase) are very undesirable
when the goal is to create a well defined coherent superposition for
subsequent use as an initial state in a certain process,
whereas such variations are less important in an atom interferometer.
Following this argument, we propose a {\it smooth}, rather than abrupt,
realization, which is advantageous in achieving
adiabaticity (and hence, robustness) more easily.
It makes use of only two laser pulses
(which can be derived from a single laser) and at the same time ensures
the automatic fulfillment of the asymptotic conditions for f-STIRAP.
Furthermore, we derive simple analytic estimates of the robustness of
f-STIRAP against variations in the pulse intensity, the pulse delay,
and the intermediate-state detuning, and discuss a possible
extension of this scheme to multistate systems.

We note that another extension of STIRAP aimed at creating a coherent
superposition of states has been proposed very recently \cite{Unanyan98}.
It is based on adiabatic transfer in a four-state system with a tripod
linkage by means of three laser pulses.

This paper is organized as follows.
The idea of f-STIRAP is presented in Sec. \ref{Sec-idea}.
The robustness of the process is examined in Sec. \ref{Sec-robustness}.
The extension of f-STIRAP to multistate systems is
discussed in Sec. \ref{Sec-multi}.
The conclusions are summarized in Sec. \ref{Sec-conclusions}.

%======================================================================

\section{Fractional STIRAP}

\label{Sec-idea}

\subsection{The idea}

The probability amplitudes $c_k(t)$ of the three states $\psi_k$ ($k=1,2,3$)
satisfy the Schr\"odinger equation,
$$
i\hbar \frac d{dt}{\bf c}(t) = \H(t){\bf c}(t), 
$$
where ${\bf c}(t)$ is a column vector comprising $c_k(t)$.
In the rotating-wave approximation,
the Hamiltonian of our three-state system is \cite{Shore90}
$$
\H(t) = \hbar \left[ 
\ba{ccc}
0 & \case12\pump(t)e^{-i\phaseP} & 0 \\ 
\case12\pump(t)e^{i\phaseP} & \Delta & \case12\Stokes(t)e^{-i\phaseS} \\
0 & \case12\Stokes(t)e^{i\phaseS} & 0
\ea
\right] , 
$$
where $\Delta $ is the single-photon detuning of the intermediate state,
while $\pump(t)$ and $\Stokes(t)$ are the Rabi frequencies
of the pump and Stokes pulses, respectively, and
$\phaseP$ and $\phaseS$ are the phases of the two fields.
An important condition for the success of the scheme discussed below is the
two-photon resonance between states $\psi_1$ and $\psi_3$,
already assumed in $\H(t)$.
The system is supposed to be initially in state $\psi_1$,
\be
\label{initial}
\Psi(-\infty) = \psi_1,
\ee
and we wish to transform it at the end of the interaction into the coherent
superposition
\be
\label{final}
\Psi(+\infty) = \psi_1\cos\mixangle - \psi_3e^{i\phase}\sin\mixangle,
\ee
where $\mixangle $ is a constant mixing angle
($0\leqq\mixangle \leqq \case12 \pi $), and $\phase$ is a constant phase.
The minus sign, which does not limit the generality of the superposition,
is taken for the sake of later convenience.

Following STIRAP, we are going to use the fact that one of the eigenvalues
of $\H(t)$ is equal to zero and the corresponding eigenstate
(the trapped state) is 
\be
\label{ATstate}
\ATS(t)=\frac 1{\OmRMS(t)}
	\left[\Stokes(t)e^{-i\phaseS}\psi_1-\pump(t)e^{i\phaseP}\psi_3\right],
\ee
where 
\be
\label{Omega}
\OmRMS(t)=\sqrt{\pump^2(t) + \Stokes^2(t)}.
\ee
In STIRAP, the Stokes pulse precedes the pump pulse,
that is $\pump(t)/\Stokes(t)\rightarrow 0$ as $t\rightarrow -\infty $ and
$\Stokes(t)/\pump(t)\rightarrow 0$ as $t\rightarrow +\infty $.
Consequently, we have $\ATS(-\infty )=e^{-i\phaseS}\psi_1$ and
$\ATS(+\infty )=-e^{i\phaseP}\psi_3$, which ensures complete
population transfer from state $\psi_1$ to state $\psi_3$
in the adiabatic limit.
Moreover, insofar as the trapped state $\ATS(t)$ does not involve
state $\psi_2$, the latter is not populated during the transfer.
This implies that its properties, including its detuning $\Delta $ and
decay rate, do not affect the transfer efficiency in the adiabatic limit.
Furthermore, since STIRAP is an adiabatic process, it is robust against
moderate variations of laser parameters (intensities, detunings,
pulse shapes, pulse widths, pulse delay).
Note that as far as STIRAP is concerned the phases of the pump and
Stokes fields do not affect the population transfer.

Let us now consider a slightly changed pulse timing, in which the Stokes
pulse still comes first and is followed after a certain time delay by the
pump pulse, but the two pulses {\it vanish simultaneously},
\be
\label{timing}
\lim _{t\rightarrow -\infty }\frac{\pump(t)}{\Stokes(t)}=0,\qquad
\lim _{t\rightarrow +\infty }\frac{\pump(t)}{\Stokes(t)}=\tan\mixangle .
\ee
In this case, the trapped state has the limits
$\ATS(-\infty )=e^{-i\phaseS}\psi_1$ and
$\ATS(+\infty )=e^{-i\phaseS}\left[ \psi_1\cos \mixangle
 - \psi_3e^{i\phase}\sin \mixangle \right]$, with
\be
\label{phase}
\phase = \phaseP + \phaseS.
\ee
Hence, if the evolution is adiabatic, the system will remain in the same
adiabatic state $\ATS(t)$ in which it is initially and, as desired,
will end up in the superposition state (\ref{final}), up to an
irrelevant common phase factor.
Moreover, since it is based on the same physical mechanism as STIRAP, f-STIRAP
should have the same properties in terms of efficiency and robustness.

The case when $\pump/\Stokes\rightarrow 1$ at $+\infty$ and $\phase=0$
is particularly interesting for quantum information and atom optics; then
$\mixangle=\case14 \pi$ and $\Psi(+\infty)=\case1{\sqrt{2}}(\psi_1-\psi_3)$.
The creation of this superposition corresponds to the Hadamard transform
of a quantum bit in quantum information.
In atom optics, if the pump and Stokes pulses propagate in opposite
directions, the creation of this coherent superposition is
accompanied with a momentum transfer of $2\hbar k$ for a half of the atoms
and f-STIRAP works in this case as a {\it coherent beam splitter}
\cite{Marte91,Lawall94BS,Goldner94,Weitz94PRA,Weitz94PRL,Weiss94,%
Featonby95,Theuer98}.

\subsection{The system}

%***************************************************************
\begin{figure}[tb]%[htbp]
%\vspace*{-90mm}
\centerline{\psfig{width=70mm,file=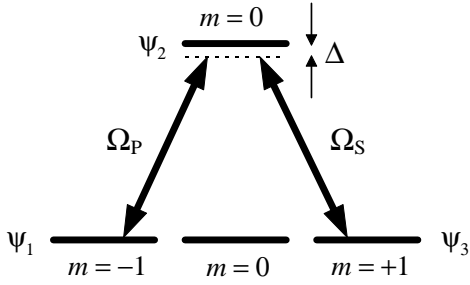}}
\vspace*{2mm}
\caption{
An example of a three-state system for realization of fractional STIRAP.
The three-state chain is formed by the sublevels in the
$J=1\leftrightarrow J=0$ transition by using a couple of circularly
polarized laser pulses.
If, for example, the system is initially in the $m=-1$ sublevel of the lower
level, then the pump pulse $\pump$ should be $\sigma ^+$ polarized
and the Stokes pulse $\Stokes$ should be $\sigma^-$ polarized.
The f-STIRAP creates a coherent superposition of states $m=-1$ and $m=+1$.
}
\label{Fig-3SS}
\end{figure}
%***************************************************************

A particularly suitable system for realization of f-STIRAP is the
three-state chain which is formed by the sublevels in
$J=1\leftrightarrow J=0$ or $J=1\leftrightarrow J=1$ transitions
by using a couple of $\sigma ^{+}$ and $\sigma ^{-}$ polarized laser pulses,
with the system prepared initially (e.\ g. by optical pumping) in the $m=-1$
(or $m=+1$) sublevel of the lower level.
This system is shown in Fig.\ \ref{Fig-3SS}.
In this case, f-STIRAP creates a coherent superposition of states $m=-1$
and $m=+1$.
If, for example, the system is initially in the $m=-1$ sublevel, then the pump
pulse should be $\sigma ^+$ polarized
and the Stokes pulse $\sigma^-$ polarized.
The convenience of this system derives from the fact that
the two-photon resonance condition, which is essential for f-STIRAP
as well as for the standard STIRAP, is automatically fulfilled,
provided there are no magnetic fields.
Moreover, the two pulses can be derived from the same laser pulse
by beam splitting.
This fact, along with the robustness of the scheme against single-photon
detuning and laser power, makes f-STIRAP insensitive to phase and energy
fluctuations.
Moreover, the equal energies of states $m=-1$ and $m=+1$ mean that
the phase difference between these two states is determined entirely
by f-STIRAP and remains fixed after its completion.
Subsequently, if necessary, the relative phase between states $\psi_1$
and $\psi_3$ can be altered by applying a pulsed magnetic field
or off-resonant laser pulses.

\subsection{The pulse sequence}

One of the possible realizations of the f-STIRAP asymptotic conditions
(\ref{timing}) is to take a Stokes pulse of longer duration than
the pump pulse, with the maxima of both pulses occuring at the same time
($t=0$), and truncate both of them there, e.\ g. 
\bmla
&&\pump(t)=\left\{ \ba{ll}
\Omeg0\sin \mixangle e^{-(t/T)^2}, & t\leqq 0 \\ 0, & t>0\ea\right. ,\\
&&\Stokes(t)=\left\{ \ba{ll}
\Omeg0\cos \mixangle e^{-(at/T)^2}, & t\leqq 0 \\ 0, & t>0\ea\right. ,
\emla
where $a<1$.
This case, a variation of which has been implemented in \cite{Weitz94PRL},
is a straightforward example of interrupted evolution.

A more elegant, smooth realization of condition (\ref{timing}) can be
achieved by using three pulses, a pump pulse and two Stokes pulses --
one with the same time dependence as the pump pulse and another
coming earlier, e.g. 
\bml
\label{smooth}
\bea
&&\pump(t)=\Omeg0\sin \mixangle e^{-(t-\tau)^2/T^2},\\
&&\Stokes(t)= \Omeg0 e^{-(t+\tau)^2/T^2}
		+ \Omeg0 \cos \mixangle e^{-(t-\tau)^2/T^2}.
\eea
\eml
In fact, this pulse timing can be achieved by using only two pulses --
one with $\sigma^-$ polarization and Rabi frequency
$\Omeg0 e^{-(t+\tau)^2/T^2}$, and another with time dependence
$\Omeg0 e^{-(t-\tau)^2/T^2}$ and elliptic polarization in the $xy$-plane,
whose electric field (in the complex representation $E=E_x+iE_y$)
is given by \cite{Shore90,Klein70}
$$
E = A_P e^{-i\omega t + i \phaseP} + A_S e^{i\omega t - i \phaseS},
$$
where $A_P/A_S = \tan\mixangle$.
The former term represents the $\sigma^+$ polarized component and
the latter term represents the $\sigma^-$ one.
Here $\case12\phase=\case12(\phaseP+\phaseS)$ is the angle of rotation
of the polarization ellipse and $|1-\tan\mixangle|/(1+\tan\mixangle)$
is its axial ratio \cite{Klein70,Suominen91}.
Thus, the desired final superposition of states (\ref{final})
is controlled entirely by the polarization of the delayed pulse.
Using a linear polarization and $\phase =0$ would lead
to $\mixangle =\case14\pi $ and a superposition
$\Psi(t\rightarrow \infty )=\case1{\sqrt{2}}(\psi_1-\psi_3)$.

Note that the phase of the first $\sigma^-$ pulse is irrelevant in the
present context [because it does not affect conditions (\ref{timing})]
and it is assumed to be the same as that of the $\sigma^-$
component of the second pulse.

\begin{figure}[tb]%[htbp]
%\vspace*{-90mm}
\centerline{\psfig{width=80mm,file=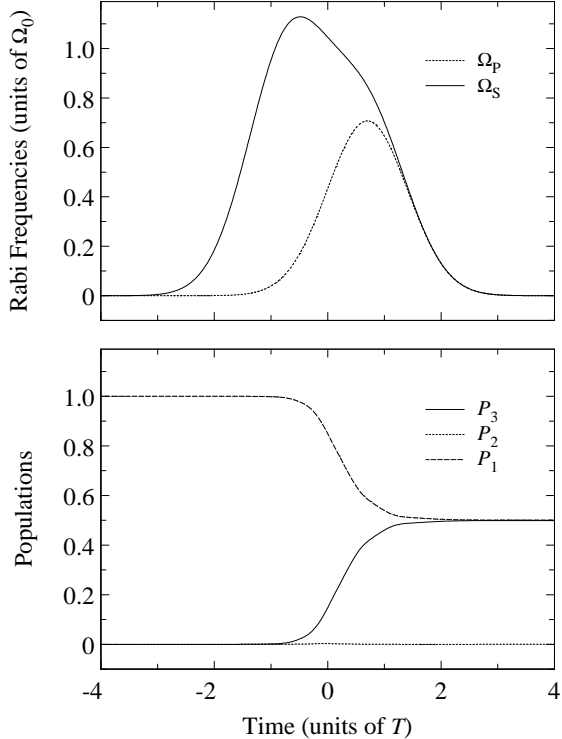}}
\vspace*{2mm}
\caption{
A typical example of time evolution (lower figure) in fractional STIRAP
in the resonance case ($\Delta=0$).
The pulse shapes (upper figure) are defined by Eqs. (\protect\ref{smooth})
with $\mixangle =\case14 \pi$, $\tau =0.7T$, $\Omeg0T=20$.
}
\label{Fig-evolution}
\end{figure}

A typical example of time evolution in fractional STIRAP is shown
in Fig.\ \ref{Fig-evolution} (lower figure).
The pulse shapes (upper figure) are defined by Eqs. (\ref{smooth}) with
$\mixangle=\case14\pi$, $\tau=0.7T$, $\Omeg0T=20$.
The population evolves smoothly from state $\psi_1$ initially to the coherent
superposition $\case1{\sqrt{2}}(\psi_1-\psi_3)$ finally, very similarly to
the manner in which the population is transferred completely from state
$\psi_1$ to state $\psi_3$ in the standard STIRAP.

%***************************************************************
%***************************************************************

%======================================================================

\section{Robustness}

\label{Sec-robustness}

\subsection{Pulse delay and laser intensity}

\subsubsection{Adiabaticity condition}

The starting point in our analysis of the robustness of f-STIRAP
against variations in the interaction parameters
is the adiabatic condition, which for $\Delta=0$ reads
\cite{Kuklinski89,Gaubatz90}
\be
\label{AdbCond}
\left| \NAC (t)\right| \ll \case12 \OmRMS(t), 
\ee
where $\th(t)=\arctan [\pump(t)/\Stokes(t)]$
and $\OmRMS(t)$ is defined by Eq. (\ref{Omega}).
For the shapes (\ref{smooth}) with $\phase=0$, we find
\bmla
&&\NAC(t)=\frac{4\tau}{T^2} \frac{\zeta(t) \sin\mixangle }
	       {\sin^2\mixangle + [\cos\mixangle + \zeta(t)]^2},\\
&&\OmRMS(t)=\Omeg0 e^{-(t-\tau)^2/T^2}
	    \sqrt{\sin^2\mixangle + [\cos\mixangle + \zeta(t)]^2}, 
\emla
where $\zeta(t)= e^{-4\tau t/T^2}$.
For appreciable non-adiabatic transitions to occur,
two conditions have to be satisfied, 
\be
\label{NAconditions}
\left| \NAC (t)\right| \gtrsim \case12 \OmRMS(t),\qquad
\left| \NAC (t)\right| \gtrsim \frac 1T. 
\ee
The former of these means that the adiabatic condition (\ref{AdbCond}) has
to be violated while the latter requires appreciable non-adiabatic coupling.
Non-adiabatic transitions are most likely to occur in the region around
the maximum of $\NAC (t)$ which is situated at
$\zeta(\Tmax)=1$, i.\ e. at 
\be
\label{tmax}
\Tmax=0.
\ee
At $t=\Tmax$, $\NAC(t)$ and $\OmRMS(t)$ are equal to 
\bmla
\label{Maxtheta}
&&\NAC_{\max}=\NAC(\Tmax)
	=\frac {2\tau} {T^2} \tan \case12\mixangle, \\
\label{MaxOmega}
&&\OmRMS(\Tmax)=2\Omeg0 e^{-\tau^2/T^2} \cos \case12\mixangle.
\emla
Note that this is not necessarily the maximum of $\OmRMS(t)$.
Since both $\OmRMS(t)$ and $\NAC(t)$ are pulse-shaped,
it is useful to find their widths (full widths at half maximum), 
\bml
\label{widths}
\bea
\label{Ttheta}
&&T_{\NAC}\approx \frac{T^2}\tau
 \ln\left(\sqrt{1 + \cos^2 \case12\mixangle}
 + \cos \case12\mixangle \right),\\
\label{TOmega}
&&T_\OmRMS\approx 2\tau +2T\sqrt{\ln 2}. 
\emla

\subsubsection{Lower limit on $\tau $}

When $\tau \rightarrow 0$, the non-adiabatic coupling $\NAC(t)$
becomes a very broad function, broader than $\OmRMS(t)$
[see Eqs. (\ref{widths})], and there are early times as well as late times,
when conditions (\ref{NAconditions}) are satisfied.
The interference between these two non-adiabatic zones leads to oscillations.
This problem will be avoided if the width of $\NAC(t)$ is smaller than
the width of $\OmRMS(t)$, $T_{\NAC}\lesssim T_\OmRMS$.
From here we find a lower bound for $\tau$, which for
$\mixangle=\case14\pi$ reads as
\be
\label{TauMinF}\tau \gtrsim 0.35T.
\ee
The same arguments applied to STIRAP lead to the estimate 
$\tau \gtrsim 0.30T$.
Hence, fractional STIRAP requires slightly larger pulse delays
than STIRAP.

\subsubsection{Upper limit on $\tau $}

Since the maximum (\ref{Maxtheta}) of the non-adiabatic coupling $\NAC(t)$
increases with $\tau$ whereas its width (\ref{Ttheta}) decreases,
$\NAC(t)$ approaches a $\delta $-function behavior for large $\tau $,
which increases the probability for non-adiabatic transitions near $\Tmax$.
The situation is aggravated by a ``hole'' in $\OmRMS(t)$, which appears
around $\Tmax$ for large $\tau$ ($\gtrsim 0.8T$).
Hence, the pulse delay should not be very large.
To suppress the non-adiabatic transitions, we must have
$\case12\OmRMS(\Tmax)\gtrsim n\NAC (\Tmax)$,
where $n$ is a ``sufficiently large'' number.
By assuming that both $\NAC(t)$ and $\OmRMS(t)$ are nearly constant near
$\Tmax$, we find that the probability for nonadiabatic transitions
in this region is $\lesssim 1/(n^2+1)$.
Hence, the choice of $n$ depends on how much nonadiabaticity we can allow, that
is how much deviation from the desired superposition is acceptable.
Thus, we find an (implicitly defined) upper bound on $\tau$,
\be
\label{TauMaxF}
\Omeg0T \gtrsim \frac{2n \sin \case12\mixangle}
	   {\cos^2 \case12\mixangle} \frac \tau T e^{\tau^2/T^2},
\ee
which is similar to the one for standard STIRAP \cite{Vitanov97D}.
This inequality can also be seen as a lower bound for the peak Rabi
frequency $\Omeg0$.
Obviously, the Rabi frequency needed to ensure sufficient adiabaticity
increases exponentially with $\tau$.

The conclusion is that although fractional STIRAP should work for any
pulse delay $\tau >0$ for sufficiently strong laser pulses, there is an
optimal range of $\tau $, in which adiabaticity is most easily achieved.
For example, for $n=5$ this range is
$0.35T\lesssim \tau \lesssim 0.93T$ for $\Omeg0T=20$,
$0.35T\lesssim \tau \lesssim 1.12T$ for $\Omeg0T=40$, and
$0.35T\lesssim \tau \lesssim 1.28T$ for $\Omeg0T=60$.

%***************************************************************
\begin{figure}[tb]
\vspace*{-2mm}
\centerline{\psfig{width=70mm,file=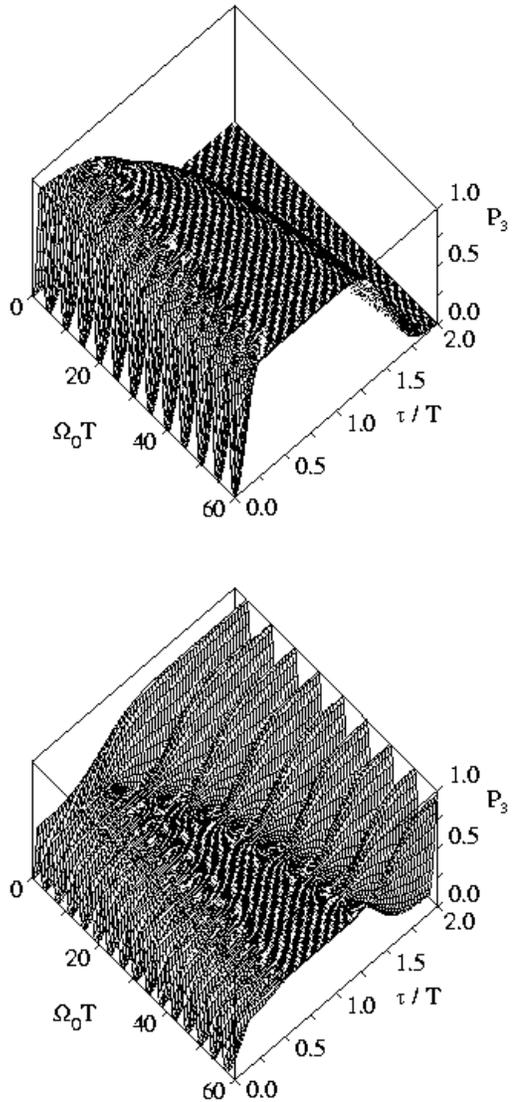}}
\vspace*{5mm}
\caption{
The final population $P_3$ of state $\psi_3$ plotted
against the time delay $\tau$ and the peak Rabi frequency $\Omeg0$
in the on-resonance case ($\Delta=0$).
The upper figure is for standard STIRAP (with $\mixangle =\case12\pi$)
and the lower figure for fractional STIRAP (with $\mixangle =\case14\pi$).
The pulse shapes are defined by Eqs. (\protect\ref{smooth}).
The plateau for STIRAP corresponds to $P_3=1$
and that for f-STIRAP to $P_3=0.5$.
}
\label{Fig-tau}
\end{figure}
%***************************************************************

In Fig.\ \ref{Fig-tau}, the final population $P_3$ of state $\psi_3$ is plotted
against the time delay $\tau$ and the peak Rabi frequency $\Omeg0$.
Comparison is made between standard STIRAP (upper figure)
and fractional STIRAP (lower figure).
The plateau for STIRAP corresponds to $P_3=1$,
while that for f-STIRAP corresponds to $P_3=0.5$.
It is seen that the two plateaus look very similar which means that
f-STIRAP should have similar properties of terms of
robustness and efficiency as STIRAP.
The borders of the plateaus are in good agreement with
our simple analytic formulas.
The oscillations (against $\Omeg0$) seen
for small $\tau $ in both plots are due to the interference between the
transitions in the two non-adiabatic regions, as discussed above.
The oscillations seen for f-STIRAP at large $\tau $ appear because then the two
components that form the Stokes pulse (the early $\sigma^-$-pulse and the
$\sigma^-$-component of the delayed elliptically polarized pulse)
are too separated and the one which comes first has almost no effect;
then the excitation dynamics is essentially the one of completely
overlapping pump and Stokes pulses \cite{Vitanov98JPB}.

\subsection{Detuning and laser intensity}

Our analysis of the robustness of f-STIRAP
against variations in the intermediate-state detuning $\Delta $
begins again from the adiabatic condition.
In this case it has the form (for $\phase=0$) \cite{Vitanov97D}
\be
\label{AdbCond-D}
n\left| \NAC (t)\right| \lesssim 
	\case12 \OmRMS(t) \frac{\sin \angleD(t)}{\cos^2\angleD(t)}, 
\ee
where $n$ is a suitably chosen large number
and the angle $\angleD $ is defined by
$$
\tan 2\angleD (t)=\frac{\OmRMS(t)}\Delta . 
$$
The results in \cite{Vitanov97D} for the detuning dependence in STIRAP suggest
that in the near-adiabatic regime (when $\Omeg0T\gg 1$), the range of
detunings which do not affect significantly the transfer efficiency
is large compared with the peak Rabi frequency, $\Delta \gg \Omeg0$.
Then $\angleD(t)\approx \OmRMS(t)/2\Delta$,
and the adiabatic condition (\ref{AdbCond-D}) reduces to
$$
\Delta \lesssim\frac{\OmRMS^2(t)}{4n\left| \NAC (t)\right| }. 
$$
It is most important to satisfy this condition in the region around
$t=0$, where $\NAC(t)$ is maximal.
There we have 
\be
\label{D-condition}
\Delta \lesssim \frac {\cos^3 \case12\mixangle}
	{2n \sin \case12\mixangle} \frac {T^2}\tau e^{-2\tau^2/T^2}
	\Omeg0^2. 
\ee
Hence, the acceptable range of intermediate detunings $\Delta$
is proportional to the squared peak Rabi frequency $\Omeg0$.

%***************************************************************
\begin{figure}[tb]%[htbp]
\vspace*{0mm}
\centerline{\psfig{width=75mm,file=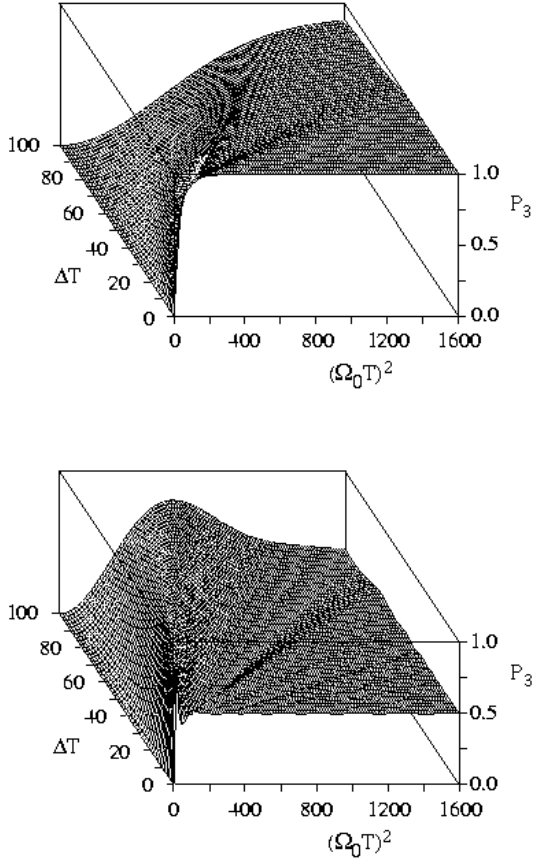}}
\vspace*{5mm}
\caption{
The final population $P_3$ of state $\psi_3$
plotted against the single-photon detuning $\Delta$
and the squared peak Rabi frequency $\Omeg0^2$.
The upper figure is for standard STIRAP (with $\mixangle =\case12\pi$)
and the lower figure for fractional STIRAP (with $\mixangle =\case14\pi$).
The pulse shapes are defined by Eqs. (\protect\ref{smooth})
with $\tau=0.7T$ in both cases.
The plateau for STIRAP corresponds to $P_3=1$
and that for f-STIRAP to $P_3=0.5$.
}
\label{Fig-Delta}
\end{figure}
%***************************************************************

In Fig.\ \ref{Fig-Delta}, the final population $P_3$ of state $\psi_3$
is plotted against the single-photon detuning $\Delta$
and the squared peak Rabi frequency $\Omeg0^2$.
Comparison is made between STIRAP (upper figure) and f-STIRAP (lower figure).
The plateau for STIRAP corresponds to $P_3=1$,
while that for f-STIRAP corresponds to $P_3=0.5$.
The plateaus are described well by our simple formula
(\ref{D-condition}).

To conclude this section, we point out that it is possible to derive
simple estimates also for the robustness of f-STIRAP
against the intermediate state loss rate in the manner
in which this has been done for STIRAP \cite{Vitanov97G}.

%======================================================================

\section{Extension to multistate systems}

\label{Sec-multi}

The $\sigma^+ \sigma^-$ pulse sequence duscussed above can easily be applied
to multistate chainwise-linked systems formed from the magnetic sublevels
in $J\leftrightarrow J^\prime = J$ or $J\leftrightarrow J^\prime = J-1$
transitions (with integer $J$), prepared initially (e.\ g., by optical pumping)
in the $m=-J$ or $m=J$ sublevel of the ground level.
The ($2J+1$)-state system formed in such a manner comprises $J+1$ sublevels
of the lower (ground) level and $J$ sublevels of the upper (excited) level.
Let us denote the amplitudes and the wavefunctions of the lower sublevels
by $c_m$ and $\psi_m$, respectively, and those of the upper sublevels
by $c_{m^\prime}^\prime$ and $\psi_{m^\prime}^\prime$
and let us assume that the atom is prepared initially in state $\psi_{-J}$.
Then the Rabi frequencies of the (up-right) transitions
$\psi_m\leftrightarrow \psi_{m+1}^\prime$ are proportional to $\pump$
times the corresponding Clebsch-Gordan coefficient
$\left( J m, 1 1 | J^\prime m^\prime \right)$ ($m^\prime=m+1$),
whereas the Rabi frequencies of the (up-left) transitions
$\psi_m\leftrightarrow \psi_{m-1}^\prime$ are proportional to $\Stokes$ times
$\left( J m, 1 (-1) | J^\prime m^\prime \right)$ ($m^\prime=m-1$),
where $\pump$ and $\Stokes$ are appropriately defined 
Rabi frequencies ``quanta''.
In the on-resonance case and in the case when only the upper sublevels
are detuned from resonance, the Hamiltonian describing this system has
a zero eigenvalue and the corresponding eigenstate has the form
\cite{Marte91,Hioe88,Shore91,Valentin94,Vitanov98}
\be
\label{ATS}
\ATS = (c_{-J},c_{-J+1}^\prime,c_{-J+2},c_{-J+3}^\prime,c_{-J+4},
	\ldots,c_{J})^T,
\ee
where the amplitudes of all upper sublevels are zero,
\be
\label{upper}
c_{-J+1}^\prime = c_{-J+3}^\prime = \ldots =c_{J-1}^\prime = 0,
\ee
while the amplitudes of the (nonzero) lower sublevels can be determined
from the recurrence relation
\be
\label{lower}
\frac{c_{m+2}}{c_{m}} =
	-\frac{\left( J m, 1 1 | J^\prime (m+1) \right)}
	      {\left( J (m+2), 1 (-1) | J^\prime (m+1) \right)}
	 \frac{\pump}{\Stokes} e^{i\phase},
\ee
with $m=-J,-J+2,-J+4,\ldots,J-2$.
The pulse sequence (\ref{timing}) in which the Stokes pulse is the first
to come ensures that state $\ATS$ is equal to state $\psi_1$ initially
and in the adiabatic limit the system stays in state $\ATS$ all the time.
Since the amplitudes (\ref{upper}) of the upper sublevels are equal to zero
these sublevels remain unpopulated, even transiently, which means that the
properties of the upper level, including decay, do not affect the process.
The selection rules leave a priori unpopulated
the other upper sublevels as well as the sublevels
$\psi_{-J+1},\psi_{-J+3},\psi_{-J+5},\ldots ,\psi_{J-1}$ of the lower level.
At the end of the pulse sequence only the sublevels
$\psi_{-J},\psi_{-J+2},\psi_{-J+4},\ldots ,\psi_{J}$
of the lower level are populated.
The created final superposition $\Psi$ depends on the ratio $\tan\mixangle$
between the pump and Stokes pulses, the phase $\phase$
and the Clebsch-Gordan coefficients characterizing the transition.

For the sake of simplicity, let us assume that $\pump/\Stokes\rightarrow 1$
as $t\rightarrow+\infty$ (i.\ e., $\alpha=\pi/4$) and $\phase=0$.
Then the superpositions $\Psi(J\leftrightarrow J^\prime)$ created by f-STIRAP
for a few most frequently used $J\leftrightarrow J^\prime$ transitions look as
\bea
&& \Psi(1\leftrightarrow 0) =
	 \sqrt{\case12}\psi_{-1}
	-\sqrt{\case12}\psi_{1},\\
&& \Psi(1\leftrightarrow 1) =
	 \sqrt{\case12}\psi_{-1}
	+\sqrt{\case12}\psi_{1},\\
&& \Psi(2\leftrightarrow 1) =
	 \sqrt{\case18}\psi_{-2}
	-\sqrt{\case34}\psi_{0}
	+\sqrt{\case18}\psi_{2},\\
&& \Psi(2\leftrightarrow 2) =
	 \sqrt{\case38}\psi_{-2}
	+\case12\psi_{0}
	+\sqrt{\case38}\psi_{2},\\
&& \Psi(3\leftrightarrow 2) =
	 \sqrt{\case1{32}}\psi_{-3}
	-\sqrt{\case{15}{32}}\psi_{-1}
	+\sqrt{\case{15}{32}}\psi_{1}\nonumber\\
&&\qquad\qquad\ 
	-\sqrt{\case1{32}}\psi_{3},\\
&& \Psi(3\leftrightarrow 3) =
	 \sqrt{\case5{16}}\psi_{-3}
	+\sqrt{\case3{16}}\psi_{-1}
	+\sqrt{\case3{16}}\psi_{1}\nonumber\\
&&\qquad\qquad\ 
	+\sqrt{\case5{16}}\psi_{3}\\
&& \Psi(4\leftrightarrow 3) =
	 \sqrt{\case1{128}}\psi_{-4}
	-\sqrt{\case7{32}}\psi_{-2}
	+\sqrt{\case{35}{64}}\psi_{0}\nonumber\\
&&\qquad\qquad\ 
	-\sqrt{\case7{32}}\psi_{2}
	+\sqrt{\case1{128}}\psi_{4},\\
&& \Psi(4\leftrightarrow 4) =
	 \sqrt{\case{35}{128}}\psi_{-4}
	+\sqrt{\case5{32}}\psi_{-2}
	+\sqrt{\case9{64}}\psi_{0}\nonumber\\
&&\qquad\qquad\ 
	+\sqrt{\case5{32}}\psi_{2}
	+\sqrt{\case{35}{128}}\psi_{4}.
\eea

Such multistate systems provide the possibility of robust creation of
well-defined radiatively stable coherent superpositions
of {\it more than two states}.
From the viewpoint of beam splitting, f-STIRAP applied to such
systems achieves coherent splitting of the initial atomic beam into $J+1$
components.
Moreover, we can alter the components in each superposition by changing
the mixing angle $\mixangle$, and we can alter their phases by changing
the phase $\phase$, i.\ e., we have two degrees of freedom.
Finally, if necessary, the populations of the intermediate
states can be removed (stored into metastable states or ionized) by
subsequent laser pulses, thus leaving the atom in a coherent superposition
of two states ($\psi_{-J}$ and $\psi_J$) with a large momentum difference.
It is seen from the superpositions listed above that, due to the
nature of the Clebsch-Gordan coefficients, this is relevant for the
transitions with $J^\prime=J$ only, because for $J^\prime = J-1$ the
coefficients of $\psi_{-J}$ and $\psi_J$ are too small.

%======================================================================

\section{Conclusions}

\label{Sec-conclusions}

We have discussed the properties of a technique which can create
any preselected coherent superposition of two states.
The technique, which we have called {\it fractional STIRAP} (f-STIRAP),
is based upon (incomplete) adiabatic population transfer between the
initial state $\psi_1$ and state $\psi_3$ through an intermediate state
$\psi_2$.
As in STIRAP, the Stokes pulse arrives before the pump pulse, but unlike
STIRAP, the two pulses vanish simultaneously.
This ``incompleted STIRAP'' evolution provides the possibility of ending
with the population residing in both states $\psi_1$ and $\psi_3$, rather
than being transferred entirely to state $\psi_3$, as in STIRAP.
We have suggested a smooth realization of f-STIRAP which requires only two
laser pulses (which can be derived from a single laser) and in the same time
ensures the automatic fulfillment of the asymptotic conditions for f-STIRAP.
Fractional STIRAP has all the properties that STIRAP has
in population transfer to a single state,
regarding robustness, efficiency, and simplicity, and can thus
be considered as its analog in creating coherent superpositions of states.
The robustness against variations in the laser parameters (such as detunings
and pulse areas) is a particularly important feature, which makes f-STIRAP
insensitive to phase and energy fluctuations of the laser.
In the frequently implemented setup,
where an atomic beam crosses two slightly displaced laser beams at right
angles, the atoms with different velocities have different interaction times
and then the robustness of the technique against the pulse area ensures that
virtually all atoms undergo the same excitation.
We have derived simple analytic estimates of the robustness of
f-STIRAP against variations in the pulse intensity, the pulse delay,
and the intermediate-state detuning, and have discussed a possible
extension of f-STIRAP to multistate systems.
In principle, the method can be implemented without significant complications
to any system where STIRAP has been done.

%======================================================================

\subsection*{Acknowledgments}

This work has been supported financially by the Academy of Finland.
We thank Prof. Klaas Bergmann for useful discussions.

%======================================================================

\end{document}